\documentclass{aastex}
\usepackage{emulateapj5,times,mathptm}
\singlespace
\slugcomment{Accepted for the publication in the {\it Astrophysical Journal 
Letters}}
\def\la{\mathrel{\hbox{\rlap{\hbox{\lower4pt\hbox{$\sim$}}}\hbox{$<$}}}}
\def\ga{\mathrel{\hbox{\rlap{\hbox{\lower4pt\hbox{$\sim$}}}\hbox{$>$}}}}

\shortauthors{Park}
\shorttitle{G292.0+1.8}

\begin{document}
\title{Nucleosynthesis in The Oxygen-Rich Supernova Remnant G292.0+1.8
from {\it Chandra} X-Ray Spectroscopy}
\author{Sangwook Park\altaffilmark{1},John P. Hughes\altaffilmark{2},
Patrick O. Slane\altaffilmark{3}, David N. Burrows\altaffilmark{1}, 
Peter W. A. Roming\altaffilmark{1}, John A. Nousek\altaffilmark{1},
and Gordon P. Garmire\altaffilmark{1}} 
\altaffiltext{1}{Department of Astronomy and Astrophysics, Pennsylvania State
University, 525 Davey Laboratory, University Park, PA. 16802; 
park@astro.psu.edu}
\altaffiltext{2}{Department of Physics and Astronomy, Rutgers University,
136 Frelinghuysen Road, Piscataway, NJ. 08854-8019}
\altaffiltext{3}{Harvard-Smithsonian Center for Astrophysics, 60 Garden Street,
Cambridge, MA. 02138}

\begin{abstract}

We continue our analysis of the Galactic oxygen-rich supernova 
remnant (SNR) G292.0+1.8, which was observed with the {\it Chandra 
X-ray Observatory}. The high angular resolution {\it Chandra} data 
resolve metal-rich ejecta knots as well as the shocked circumstellar 
medium. X-ray emission from the ejecta material in G292.0+1.8 is 
dominated by highly ionized O, Ne and Mg. Measured abundance ratios
suggest that this material was produced during the hydrostatic 
evolution of the massive progenitor star. In contrast to Cassiopeia A,
there is little evidence for X-ray-emitting ejecta from explosive
nucleosynthesis, i.e., material enriched in Si, S, and particularly, 
Fe. This limits the amount of mixing or overturning 
of deep ejecta material in G292.0+1.8 and suggests that the ejecta 
are strongly stratified by composition and that the reverse shock
has not propagated to the Si/S or Fe-rich zones. On the other hand, 
the bright equatorial belt is dominated by X-ray emission with normal 
chemical composition, which supports shocked dense circumstellar material 
for its origin.  We find that the thermal pressure in the SNR is much 
higher than the pressure in the pulsar wind nebula (PWN), indicating 
that the reverse shock has not yet begun to interact with the PWN.

\end{abstract}


\section {\label {sec:intro} INTRODUCTION}

The Galactic supernova remnant (SNR) G292.0+1.8 (MSH 11$-$54) is a 
source of strong oxygen and neon lines in the optical band and 
has been classified as an ``oxygen-rich'' SNR \citep{goss79,murdin79}. 
This abundance pattern is indicative of a core-collapse supernova (SN) 
from a massive progenitor. The small number of young O-rich SNRs, where 
the ejecta are not yet fully mixed with the ambient interstellar medium 
(ISM), may thus provide a rare opportunity for a detailed study of the 
SNR nucleosynthesis. For example, Hughes et al.\ (2000) have successfully 
performed such a study with a Galactic O-rich SNR Cassiopeia A, which 
revealed that explosive O- and Si-burning products from the deep interior 
of the massive progenitor dominate ejecta material in that SNR. 

The unprecedented high angular resolution image of G292.0+1.8 obtained
by the Advanced CCD Imaging Spectrometer (ACIS) on board the {\it 
Chandra X-Ray Observatory} has resolved the textbook-like structures 
in the SNR: the associated pulsar and its wind nebula (PWN), the 
metal-rich ejecta knots, the shocked circumstellar medium (CSM), and 
the blast wave shock front propagating into the ambient ISM (Hughes 
et al. 2001; Park et al. 2002, P02 hereafter). 
Earlier works (Hughes \& Singh 1994; P02; Gonzalez \& Safi-Harb 
2003, G03 hereafter) have shown that G292.0+1.8 is dominated by O, Ne, 
and Mg line emission and identified a plausible mass for the progenitor 
star of $\sim$25 M$_{\odot}$$-$40 M$_{\odot}$ based on either the 
integrated X-ray spectrum or spectra extracted over large emission 
regions. As a continuation of our previous image analysis (P02), we here 
report the results from our spectral analysis of G292.0+1.8. We take a 
different approach than the previous work by G03, in order to study 
several individual small-scale emission features which we demonstrate 
to be knots of pure or nearly pure SN ejecta. Comparisons of the 
abundance patterns in these knots to nucleosynthesis models allow us 
to trace their origin back to specific burning sites in the progenitor 
star, rather than global abundances averaged over the entire SNR. 
Although our investigation does not exhaustively present all emission 
features in the SNR, we do select a representative range of features 
based on the emission line equivalent width (EW) maps (P02). 
A description of the observation and the data reduction of the ACIS 
data can be found in P02. 

\section{\label{sec:spec} Analysis}

We corrected the ACIS data for charge transfer inefficiency
(CTI; Townsley et al. [2000; 2002a]) and use detector response 
matrices generated by Townsley et al. (2002b) for the 
CTI-corrected data. The low energy ($E$ $\la$ 1 keV) quantum 
efficiency (QE) of the ACIS has degraded because of molecular 
contamination on the optical blocking filter. We corrected 
this time-dependent QE degradation by modifying the ancillary 
response function for each extracted spectrum, utilizing the 
IDL ACISABS software\footnote{For the discussion on this 
instrumental issue, see
http://cxc.harvard.edu/cal/Acis/Cal\_prods/qeDeg/index.html.
The software was developed by George Chartas and is available at
http://www.astro.psu.edu/users/chartas/xcontdir/xcont.html.}. 

\begin{figure}[]
\figurenum{1}
\centerline{\includegraphics[angle=0,width=12cm]{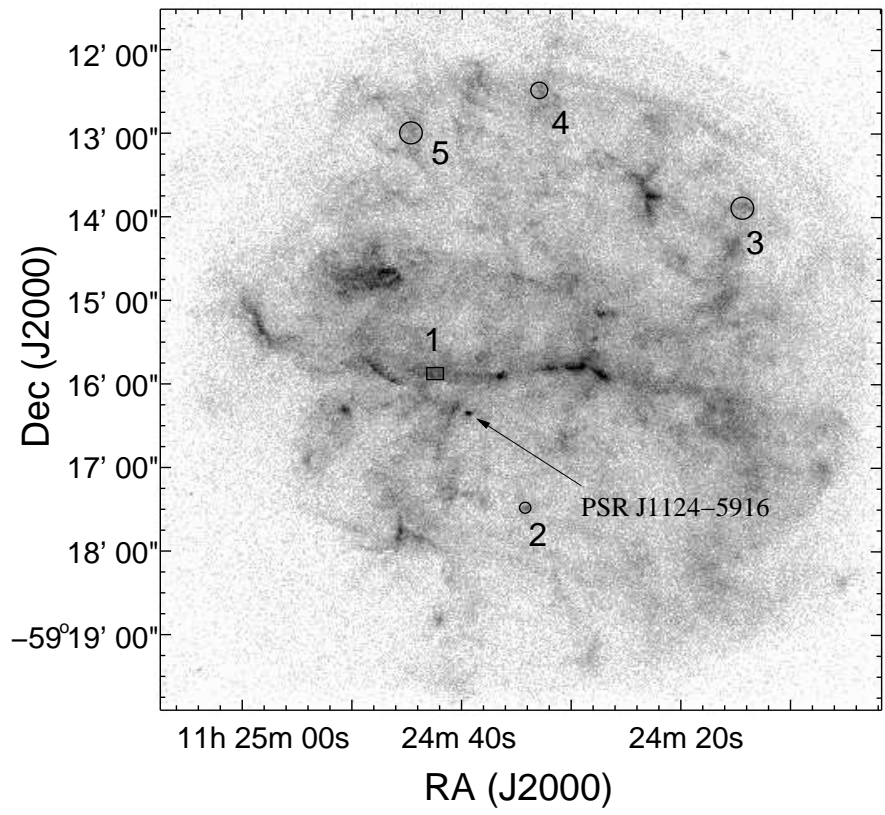}}
\figcaption[]{The gray-scale (darker scales mean brighter intensity) 
0.3 $-$ 8 keV band {\it Chandra}/ACIS image of G292.0+1.8. Regions 
1 $-$ 5 are marked. 
\label{fig:fig1}}
\end{figure}

Based on the preliminary results from the color image and EW
analysis by P02, we perform spectral 
analysis of small angular size features (typically 
$\sim$10$^{\prime\prime}$ scales)
across the SNR. We find that several regional spectra 
(Figure~\ref{fig:fig1}) represent the range of spectral 
characteristics seen in the SNR (P02). These selected regions are 
also morphologically well-defined in both the color image and the EW 
images and contain comparable photon statistics.
We subtracted background spectra from source free regions of 
the detector. Each spectrum was binned to contain at 
least 20 counts per channel and fitted in the 0.5 $-$ 4 keV band. 
We fixed the abundances for elemental species He, C, N, and Ni at 
solar \citep{anders89}, because contributions from these elements 
to the X-ray spectrum in the fitted bandpass are small.
We allow other elemental abundances to vary freely. 
In the spectral fits described below, we use a non-equilibrium
ionization (NEI) plane-parallel shock model \citep{borkowski01} 
unless otherwise noted. 

\begin{figure}[]
\figurenum{2}
\centerline{{\includegraphics[angle=0,width=10cm]{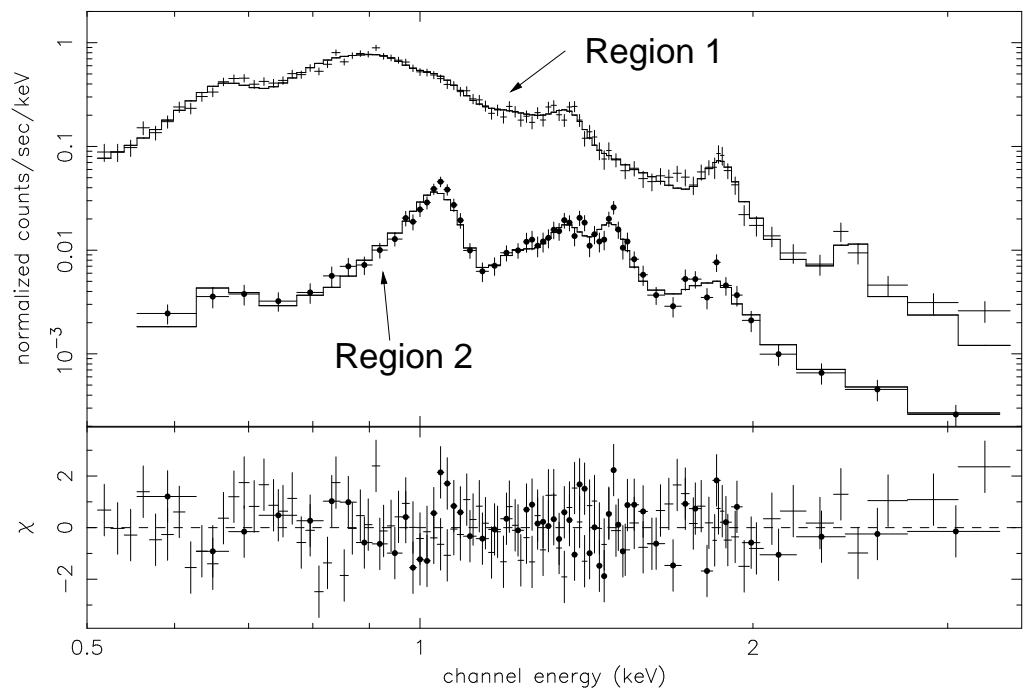}}}
\figcaption[]{Spectra from the equatorial belt (region 1) and 
the H-like Ne-enriched knot (region 2) within G292.0+1.8. Each
spectrum has been arbitrarily scaled for the purpose of display.
\label{fig:fig2}}
\end{figure}

The bright equatorial belt of G292.0+1.8 appears to be dominated by 
X-ray emission from dense CSM produced by stellar winds from 
the massive progenitor as heated by the blast wave (P02). 
A typical spectrum of this bright central belt-like feature is
represented by that of region 1. This regional spectrum contains
$\sim$8500 photons and is best fitted with an NEI shock with electron
temperature $kT$ $\sim$ 0.7 keV 
(Figure \ref{fig:fig2}; Table \ref{tbl:tab1}). The near-solar fitted
abundances (Table \ref{tbl:tab1}) are consistent with the proposed 
circumstellar origin, rather than metal-rich ejecta.

On the other hand, the spectrum from region 2 shows remarkably 
different features (Figure \ref{fig:fig2}). The region 2 spectrum
contains $\sim$2000 photons, and is dominated by emission from 
the highly ionized Ly$\alpha$ lines of Ne ($E$ $\sim$ 1.05 keV) and 
Mg ($E$ $\sim$ 1.48 keV), which unambiguously indicate emission 
from metal-rich ejecta. The strong Ne Ly$\alpha$ and Mg Ly$\alpha$ 
line emission indicates a high electron temperature and/or an advanced 
ionization state. In fact, the spectrum of region 2 can be described 
by a thermal plasma model in collisional ionization 
equilibrium (CIE) with an electron temperature of $kT$ $\sim$ 0.8 
keV. An NEI model can also fit the data with a higher electron 
temperature ($kT$ $\sim$ 5 keV; $n_et$ $\sim$ 1.5 $\times$ 10$^{11}$ 
cm$^{-3}$ s). With the current photon statistics, we cannot 
discriminate between these models. Nonetheless, strong enhancements 
in the O, Ne, and Mg abundances are present in either case. 
Although the statistical uncertainties are relatively large, O, Ne, 
and Mg abundances are evidently enhanced whereas Si, S, and Fe 
abundances are low (Table~\ref{tbl:tab1}). The well-defined, compact 
morphology of this feature (angular size of $\sim$8$^{\prime\prime}$ 
in diameter, physical size of $\sim$0.2 pc for a distance of $d$ = 6 
kpc for G292.0+1.8 as recently determined by Gaensler \& Wallace 
[2003]) allows us to employ a local background subtraction. We then
find that this regional spectrum can be described with emission 
from only O, Ne, Mg, and Si with no contribution from underlying 
continuum other than that contributed by these species. This is 
likely a knot of pure metal ejecta. 

\begin{figure}[]
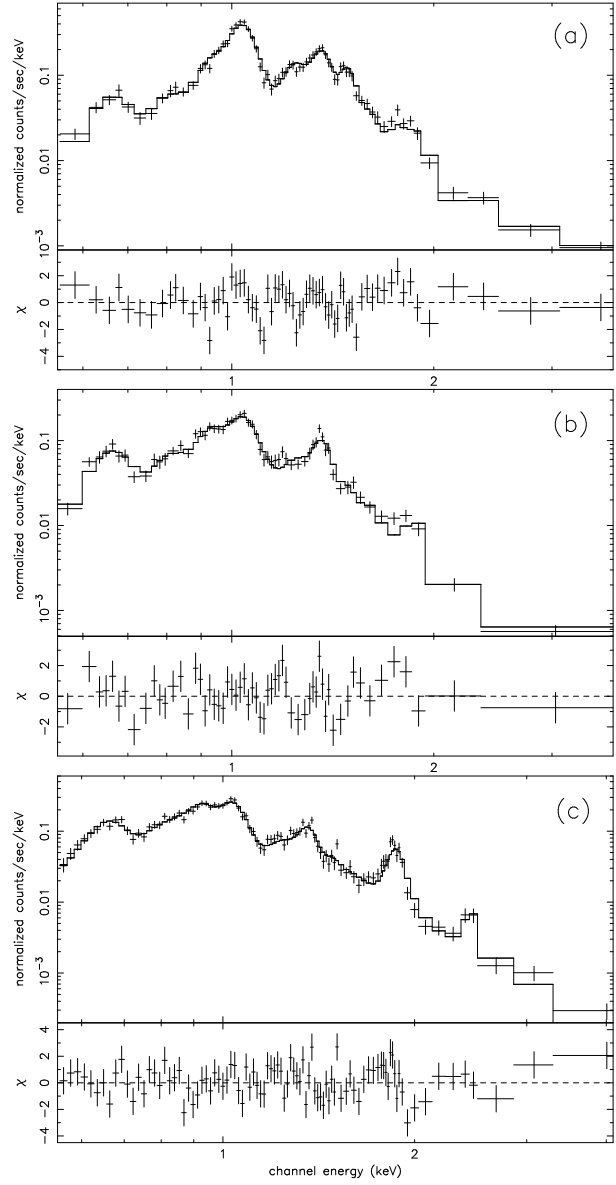

\figurenum{3}
\centerline{{\includegraphics[angle=-90,width=8cm]{f3a.ps}}}
\centerline{{\includegraphics[angle=-90,width=8cm]{f3b.ps}}}
\centerline{{\includegraphics[angle=-90,width=8cm]{f3c.ps}}}
\figcaption[]{Spectra from metal-rich knots within G292.0+1.8.
Panels (a), (b), and (c) are spectra from regions 3, 4, and 5,
respectively.
\label{fig:fig3}}
\end{figure}

The northern boundary of the SNR has a few knots of emission which 
appear to be clumpy ejecta material embedded in fainter diffuse
emission (see Figure~1 in P02). These features are also strong in 
the EW images (P02). Spectra from these knots can be 
characteristically represented by those from regions 3, 4, and 5
(Figure~\ref{fig:fig1}). Regions 3 and 4 contain $\sim$6200 
and $\sim$3700 photons, respectively, and are best fitted with a high
temperature ($kT$ $\sim$ 3 $-$ 5 keV) NEI plasma 
(Figure~\ref{fig:fig3}a; Figure~\ref{fig:fig3}b; Table~\ref{tbl:tab1}). 
The metal abundances are highly enhanced, confirming that these knots are 
ejecta. Table~\ref{tbl:tab2} presents the abundances of O, Ne, and 
Mg relative to Si; these abundance ratios are much higher than
solar ratios and are comparable to those in region 2. 
The region 5 spectrum is described with a low electron temperature 
of $kT$ $\sim$ 0.6 keV (Figure~\ref{fig:fig3}c; Table~\ref{tbl:tab1}), 
which is similar to that of region 1. The ionization timescale 
is significantly larger than those for regions 3 and 4. The best-fit 
abundances for all fitted elements are a few times higher than solar, 
but, unlike regions 2, 3, and 4, the abundance ratios with respect to
Si are less enhanced (Table~\ref{tbl:tab2}). Region 5 thus appears to be 
ejecta material that is relatively more enhanced in Si and S than other 
portions of the SNR.

\section{\label{sec:disc} DISCUSSION}

X-ray spectra from regions 2, 3, and 4 unambiguously demonstrate 
characteristics of metal-rich ejecta, which are dominated by
emission from highly ionized O, Ne, and Mg. Low-Z elements
typically dominate the ejecta throughout the SNR,
and we find little evidence of enriched heavier elements such as Si,
S, and Fe. This is in striking contrast to the features discovered
in Cas A in which strongly enhanced Si, S, and Fe abundances are
present \citep{hughes00}. We have found only marginal evidence of
Si- and S-rich ejecta in G292.0+1.8 (e.g., region 5). In core-collapse
nucleosynthesis models, low-Z elements are primarily produced
in the outer layers of the unprocessed (i.e., hydrostatic He-burning)
C-core while high-Z species are synthesized by explosive Ne-, O- and/or
Si-burning in the deep interior of the massive progenitor star
(e.g., Thielemann et al. 1996). Our results thus suggest that 
extensive mixing or overturning of the explosive nucleosynthesis products
has not occurred in G292.0+1.8 and furthermore that the reverse shock
in the remnant has not propagated to the Si/S or Fe-rich zones.
The alternate possibility, that no Fe or Si/S ejecta were produced, seems
unlikely.

Ne appears to be considerably more abundant in G292.0+1.8 
than predicted by standard 
core-collapse models in which O is invariably 
overabundant relative to Ne. For instance, assuming 
$n_e$ $\approx$ 9$n_{Ne}$ (where $n_e$ is the electron density 
and $n_{Ne}$ is the Ne ion density) for the mean ionization state 
implied by the observed spectrum of region 2, $n_e$ $\sim$ 2 
cm$^{-3}$ and so a Ne ejecta mass of $M_{Ne}$ $\sim$ 8 $\times$ 
10$^{-4}$ $M_{\odot}$ can be derived (we assumed a spherical 
geometry with an apparent angular radius of 4$^{\prime\prime}$ 
and a distance of 6 kpc). Similarily, an O ejecta mass of $M_{O}$ 
$\sim$ 11 $\times$ 10$^{-4}$ $M_{\odot}$ is derived. This indicates
that the O and Ne ejecta masses are nearly comparable, whereas models
typically predict a several times larger mass of O than that of Ne
(e.g., Thielemann et al. 1996). This Ne overabundance is thus 
difficult to understand in terms of standard core-collapse 
nucleosynthesis models. An extensive study of nucleosynthesis 
models for various conditions of stellar structure and burning 
processes may be necessary in order to interpret this 
anomaly, which is beyond the scope of this {\it Letter}.

We compared the measured abundance ratios of O:Si, Ne:Si, and Mg:Si 
with available core-collapse nucleosynthesis models. 
The O:Si abundance ratios (in number) are larger than solar by a 
factor of $\ga$5 for regions 2, 3, and 4. Ne:Si and Mg:Si ratios
are also large, $\sim$4 $-$ 15 times higher than solar. Assuming
$^{16}$O, $^{20}$Ne, $^{24}$Mg, and $^{28}$Si to be the dominant 
isotopes, we derive elemental mass ratios in
each region (Table~\ref{tbl:tab2}). We compare these mass ratios
with the mass fraction ratios from model nucleosynthesis of specific
burning sites interior to a 20 $M_{\odot}$ \citep{thielemann96} and
a 25 $M_{\odot}$ \citep{woosley95} progenitor. The observed O:Si, 
Ne:Si, and Mg:Si mass ratios for regions 2, 3, and 4 appear to be 
roughly consistent with those of nucleosynthesis model products of 
unprocessed C-core material from the hydrostatic evolution
of the massive progenitor and are not a good match to abundances 
from the explosive Ne-burning regardless of the nucleosynthesis
models (Table~\ref{tbl:tab2}). With only the abundances of 
hydrostatic burning products available, it is difficult to constrain 
the mass of the progenitor star. Reliable progenitor masses may be 
determined by comparing both the hydrostatic and the explosive 
nucleosynthesis products between the observations and the models. 
Previous studies of G292.0+1.8 (e.g., Hughes \& Singh 1994; G03) 
derived progenitor masses by assuming that the entire ejecta was 
emitting X-rays so that the measured abundances could be directly 
compared to integrated model yields. This assumption now appears to 
be invalid based on our results.

We estimate the volume of region 1 using a simple slab-like 
cylindrical geometry with an apparent angular size
of 11$\farcs$6 $\times$ 8$\farcs$6  (0.34 $\times$ 0.25 pc) and a
depth of $\sim$0.15 pc along the line of sight. This, plus the
overall spectral normalization, results in a value for the
postshock electron density of $n_e$ $\sim$ 40 cm$^{-3}$. 
Similarly high density values are found among the other
bright filaments of the equatorial belt. These density values are
larger than the ambient density of the blast wave 
measured at the outer boundary of the SNR ($n_0$ $\sim$ 0.2 $-$ 0.5 
cm$^{-3}$; G03), by nearly two orders of magnitudes. These results 
are supportive of the proposed dense, asymetric CSM origin for the 
equatorial belt of G292.0+1.8. Our results indicate a high 
thermal pressure ($P$ $\sim$ 8 $\times$ 10$^{-8}$ ergs cm$^{-3}$)
for the shocked CSM which is more than an order of magnitude higher
than that of the PWN ($P$ $\ga$ 10$^{-9}$ ergs cm$^{-3}$; Hughes et al.
2003). The large difference in pressures gives strong support
to the idea that the reverse shock has not reached the PWN, further
reinforcing our earlier suggestion that the reverse shock has not
propagated into the heavy element dominated central portions of the ejecta.

We conclude by summarizing our view of the nature and evolutionary
state of G292.0+1.8.  We find that the compact ejecta knots in the SNR
are dominated by the species O, Ne, and Mg, and show a wide range of
temperatures and ionization timescales. The relative abundances are
best described by the hydrostatic burning products from a massive
progenitor, although Ne appears to be relatively more
enhanced than the nucleosynthesis models predict.  There are a number
of bright dense filaments of normal solar-type composition near the
remnant's center as well as fainter, lower density shocked matter near
the outer edge. The former are believed to represent dense stellar wind
material, while the latter is probably the ambient interstellar
medium. The wide variety of densities and temperatures, as our
analysis reveals, indicates that the two-temperature picture
of G03 for G292.0+1.8 is grossly oversimplified. From our spectral
fits we measure thermal pressures in the shocked gas that are much
higher than the pressure in the PWN. This argues that
the reverse shock in G292.0+1.8 has not yet propagated deeply enough
into the ejecta to encounter the PWN. Thus there should be a reservoir
of cold unshocked ejecta between the PWN and the reverse shock. It is
plausible to conclude that this is where the Si/S and Fe rich ejecta
are in G292.0+1.8, since these species are missing in the optical and
X-ray--emitting ejecta.  In contrast to Cas A, the ejecta in G292.0+1.8
appear not to have undergone extensive mixing and overturning, but have
largely retained their original compositional stratification.

\acknowledgments
{This work was funded by NASA under contract NAS8-38252 to Penn State,
NAS8-39073 to CfA and {\it Chandra} grant G01-2052X to Rutgers University.
}

\clearpage

\begin{deluxetable}{ccccccccccc}
\footnotesize
\tablecaption{Results from Spectral Fittings\tablenotemark{a}
\label{tbl:tab1}}
\tablewidth{0pt}
\tablehead{ \colhead{Region} & \colhead{$kT$} & \colhead{$N_H$} & 
\colhead{$n_et$\tablenotemark{b}} & \colhead{O} & \colhead{Ne} & 
\colhead{Mg} & \colhead{Si}& \colhead{S} & \colhead{Fe} & 
\colhead{${\chi}^{2}/{\nu}$}\\
 & \colhead{(keV)} & \colhead{(10$^{21}$ cm$^{-2}$)} & 
\colhead{(10$^{11}$ cm$^{-3}$ s)} & & & & & & &}
\startdata
\vspace{1mm}
1 & 0.66$^{+0.12}_{-0.04}$ & 3.8$^{+1.2}_{-0.7}$ &
7.2$^{+2.5}_{-1.7}$ & 1.1$^{+0.1}_{-0.1}$ & 0.6$^{+0.2}_{-0.1}$ &
0.4$^{+0.1}_{-0.1}$ & 0.4$^{+0.1}_{-0.1}$ & 1.0$^{+0.2}_{-0.3}$ &
0.4$^{+0.1}_{-0.1}$ & 93.1/85\\
\vspace{1mm}
2\tablenotemark{c} & 0.82$^{+0.07}_{-0.07}$ & 4.4$^{+0.8}_{-1.6}$ &
- & 4.7$^{+4.1}_{-2.2}$ & 15.4$^{+11.5}_{-5.8}$ &
8.1$^{+6.0}_{-3.0}$ & 1.0$^{+0.9}_{-0.4}$ & $<$ 0.6 &
$<$ 0.1 & 61.7/51\\
\vspace{1mm}
3 & 5.31$^{+3.83}_{-2.08}$ & 7.7$^{+0.5}_{-0.5}$ &
0.8$^{+0.1}_{-0.1}$ & 57$^{+168}_{-23}$ & 129$^{+301}_{-50}$ &
72$^{+168}_{-29}$& 8.9$^{+12.1}_{-3.8}$ & 5.1$^{+2.0}_{-2.0}$ &
$<$ 1.0 & 90.8/61\\
\vspace{1mm}
4 & 3.09$^{+6.01}_{-1.28}$ & 6.4$^{+0.8}_{-1.1}$ &
0.5$^{+0.1}_{-0.1}$ & 13.0$^{+7.9}_{-2.8}$ & 21.9$^{+7.7}_{-4.9}$ &
12.4$^{+8.2}_{-2.8}$ & 2.4$^{+1.5}_{-0.6}$ & $<$ 0.6 &
2.2$^{+1.5}_{-0.7}$ & 76.3/50\\
\vspace{1mm}
5 & 0.59$^{+0.06}_{-0.04}$ & 3.0$^{+0.5}_{-0.5}$ &
80.1$^{+79.9}_{-28.1}$ & 5.3$^{+4.4}_{-1.4}$ & 7.0$^{+5.6}_{-1.7}$ &
2.8$^{+2.2}_{-0.7}$ & 4.0$^{+2.9}_{-0.9}$ & 5.5$^{+2.1}_{-1.3}$ &
0.4$^{+0.2}_{-0.1}$ & 124.4/87\\
\enddata
\tablenotetext{a}{1$\sigma$ errors are quoted for the abundances, 
otherwise, 2$\sigma$ errors are presented.}
\tablenotetext{b}{The errors have been obtained after fixing $N_H$ and 
$kT$ at the best-fit values.}
\tablenotetext{c}{Best-fit parameters from a CIE model is presented
for the region 2 spectrum.}

\end{deluxetable}

\begin{deluxetable}{cccccccccc}
\footnotesize
\tablecaption{Mass Ratios\tablenotemark{a}
\label{tbl:tab2}}
\tablewidth{0pt}
\tablehead{ \colhead{Elements} & \colhead{Region} & \colhead{Region} &
\colhead{Region} & \colhead{Region} & \colhead{C-Core\tablenotemark{b}} & 
\colhead{C-Core} & \colhead{Ex Ne\tablenotemark{b}}& \colhead{Ex Ne} & 
\colhead{Solar\tablenotemark{b}} \\
 & \colhead{2} & \colhead{3} & \colhead{4} & \colhead{5} & 20 $M_{\odot}$ 
& 25 $M_{\odot}$ & 20 $M_{\odot}$ & 25 $M_{\odot}$ & }
\startdata
\vspace{1mm}
O/Si & 64.4$^{+24.7}_{-19.2}$ & 88.2$^{+12.4}_{-10.0}$ & 
  74.2$^{+12.6}_{-8.6}$ & 18.2$^{+2.1}_{-1.4}$ & 36 & 110 & 10 & 2 & 13.7 \\
\vspace{1mm}
Ne/Si & 38.1$^{+11.4}_{-8.2}$ & 36.1$^{+4.7}_{-3.6}$ &
  22.6$^{+4.0}_{-2.9}$ & 4.4$^{+0.4}_{-0.4}$ & 6.5 & 27 & 0.5 & 0.01 & 2.5 \\
\vspace{1mm}
Mg/Si & 7.5$^{+2.2}_{-1.6}$ & 7.4$^{+1.0}_{-0.8}$ &
  4.7$^{+0.8}_{-0.6}$ & 0.6$^{+0.1}_{-0.1}$ & 4.5 & 5.3 & 1 & 0.002 & 0.9 \\
\enddata

\tablenotetext{a}{The mass ratios and their uncertainties (1$\sigma$) 
are derived from the abundance ratios with the Si abundances fixed at 
the best-fit values.}

\tablenotetext{b}{Nucleosynthesis models are based on Thielemann et al.
(1996) for the 20 $M_{\odot}$, Woosley \& Weaver (1995) for the 25 
$M_{\odot}$ progenitors. C-Core indicates abundances for the carbon-core 
region; Ex Ne indicates abundances from explosive Ne burning. 
The solar ratios are from Anders \& Grevesse 
(1989).}

\end{deluxetable}

\end{document}